\documentclass[twocolumn,showpacs,preprintnumbers,amsmath,amssymb,prb,aps]{revtex4}

\usepackage{epsfig}
\usepackage{graphicx}
\usepackage{dcolumn}
\usepackage{bm}

\begin{document}

\title{Study of magnetic interactions in a spin liquid, Sr$_3$NiPtO$_6$
using density functional approach}

\author{Sudhir K. Pandey and Kalobaran Maiti}

\altaffiliation{Electronic mail: kbmaiti@tifr.res.in}

\affiliation{Department of Condensed Matter Physics and Materials
Science, Tata Institute of Fundamental Research, Homi Bhabha Road,
Colaba, Mumbai - 400 005, India}

\date{\today}

\begin{abstract}
We investigate the magnetic interactions in Sr$_3$NiPtO$_6$,
characterized to be a spin liquid using \emph{ab initio}
calculations. The results reveal a novel metal to insulator
transition due to finite exchange interaction strength; the magnetic
solutions (independent of magnetic ordering) are large band gap
insulators, while the non-magnetic solution is metallic. The Ni
moment is found to be large and the coupling among intra-chain Ni
moments is antiferromagnetic unlike other compounds in this family.
These results, thus, reveal the importance of intra-chain
antiferromagnetic interaction in addition to geometrical frustration
to derive spin liquid phase.
\end{abstract}

\pacs{75.30.Et, 75.50.Mm, 71.70.Gm, 75.20.Hr}

\maketitle


Geometrical frustration in antiferromagnetically coupled systems
often prevents long range order. Here, the magnetic moments are
strongly correlated, still they remain paramagnetic even at zero
temperature. Such systems are called {\it spin liquids}. Certain
crystal structures e.g. Kagom$\acute{e}$ lattice, pyrochlores {\it
etc.} consisting of triangular units are frustrated and are
favorable to generate this phase. However, spin liquid phase is
barely observed experimentally as the long range order can be
achieved in such frustrated systems via small perturbations due to
disorder, pressure, application of external fields etc.

Recently, a new class of compounds, $AE_3MM^\prime$O$_6$ ($AE$ =
alkaline earths, $M$ and $M^\prime$ are transition metals) is found
having geometrical frustration. The crystal structure is
rhombohedral K$_4$CdCl$_6$ type (space group $R\overline{3}c$) as
shown in Fig. 1. It contains one-dimensional (1D) chains along
$c$-axis consisting of alternating $M$O$_6$ trigonal prisms and
$M^\prime$O$_6$ octahedra connected by face sharing. In the
$ab$-plane, the chains form a triangular lattice. Thus, these
systems exhibit fascinating electronic and magnetic properties
characteristic of 1D chains as well as those due to geometrical
frustration in the $ab$-plane. Many compounds in this class are
synthesized and studied extensively due to the finding of varieties
of interesting properties involving geometrical
frustration\cite{niharika,claridge,wuprl,wuprb,flahaut,sudhir,sampath,agrestini,niitakaprl,niitakajssc,stitzer,sengupta,whangbo,vidya,sampath1,fresard,takubo,sugiyama}
unlike other quasi-one dimensional systems having no
frustration.\cite{cowire,ybni,sr2cuo3,dimen}

\begin{figure}
\vspace{-12ex}
\begin{center}
\includegraphics[angle=0,width=0.4\textwidth]{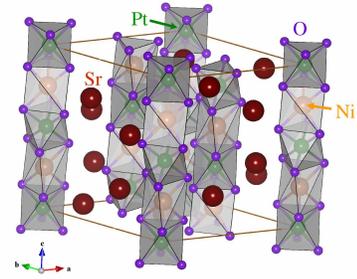}
\vspace{-36ex}
\end{center}
\caption{(Color online) The crystal structure of Sr$_3$NiPtO$_6$.
Quasi-one dimensional chains are shown by shaded regions.
 }
 \vspace{-2ex}
\end{figure}

Interestingly, only Sr$_3$NiPtO$_6$ exhibit spin liquid
behavior.\cite{niharika} Analogous compounds
Sr$_3$NiRhO$_6$\cite{niharika} and Sr$_3$CuPtO$_6$\cite{claridge}
exhibit long range order as the other compounds in this family
does.\cite{wuprl,wuprb,flahaut,sudhir} In these later systems,
geometrical frustration is manifested as partially disordered
antiferromagnetic phase, where two thirds of the chains are
antiferromagnetically coupled and the third one is incoherent.
Several contrasting suggestions exists in the literature to explain
the absence of magnetic order in Sr$_3$NiPtO$_6$. For example,
comparison of Sr$_3$NiPtO$_6$ and Sr$_3$CuPtO$_6$ indicated that
Ni$^{2+}$ could be in a singlet state.\cite{claridge} However,
Ni$^{2+}$ in a similar compound, Sr$_3$NiRhO$_6$ possess large
magnetic moment ($S$ = 1 state).\cite{niharika,sudhir} Thus, the
absence of spin liquid phase in these frustrated systems as well as
its anomaly in Sr$_3$NiPtO$_6$ is puzzling. Here, we show that Ni
moment is large and couple antiferromagnetically along the chain
unlike other compounds.\cite{niharika,sudhir,ca3co2o6} This is
probably important to derive spin liquid phase in this compound. In
addition, we observe a novel magnetization induced metal to
insulator transition.

The nonmagnetic and magnetic GGA (generalized gradient
approximation) electronic structure calculations were carried out
using {\it state-of-the art} full potential linearized augmented
plane wave (FPLAPW) method.\cite{blaha} The spin orbit coupling for
Ni and Pt were considered in the calculations. The lattice
parameters and atomic positions used in the calculations are taken
from the literature.\cite{claridge} The Muffin-Tin sphere radii were
chosen to be 2.33, 2.19, 2.01, and 1.78 a.u. for Sr, Ni, Pt, and O,
respectively. For the exchange correlation functional, we have
adopted recently developed GGA form by Wu {\em et al.}\cite{wu} The
convergence was achieved by considering 512 $k$ points within the
first Brillouin zone and the error bar for the energy convergence
was set to be smaller than 10$^{-4}$ Rydberg/cell.


\begin{figure}
\vspace{-2ex}
\begin{center}
\includegraphics[angle=0,width=0.4\textwidth]{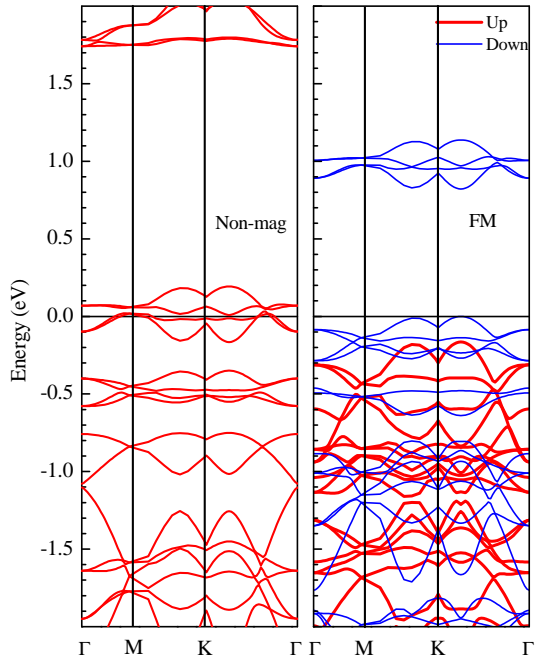}
\vspace{-4ex}
\end{center}
\caption{(Color online) Band dispersions along different symmetry
directions of the first Brillouin zone for nonmagnetic and
ferromagnetic solutions. Fermi level is denoted by zero in the
energy scale.
 }
 \vspace{-2ex}
\end{figure}

Calculated energy bands are shown in Fig. 2; the left and right
panels show the band dispersions corresponding to non-magnetic and
ferromagnetic solutions, respectively. The non-magnetic solution
clearly correspond to a metallic ground state with large
contribution at the Fermi level, $\epsilon_F$ represented by 'zero'
in the energy scale. There are four bands lying within $\pm$0.1 eV
of $\epsilon_F$. At the $\Gamma$ and $M$ points, the bands just
above and bellow the $\epsilon_F$ are two fold degenerate. The
separation of these bands are largest at $\Gamma$ point ($\sim$ 0.16
eV) and lowest at $M$ point ($\sim$ 0.04 eV). The degeneracy of
these bands is lifted along $MK$ direction.

In the ferromagnetic solution, the up-spin energy bands move away
from the Fermi level, towards lower energies. The Fermi level
appears at the top of the down spin bands. The energy gap in the
down spin channel is about 0.8 eV, which is much smaller than the
gap of$\sim$~2.5 eV in the up spin channel. Thus, in this phase, the
electronic conduction will be spin-polarized even in the insulating
phase in a large temperature range. The insulating gap of about 0.8
eV appears along the MK and $\Gamma K$ directions. The band gaps at
$\Gamma$, $K$ and $M$ points are found to be about 0.96, 1 and 1.1
eV, respectively.

The most notable observation here is the manifestation of insulating
phase due to magnetization in a non-magnetic metal. It is observed
that antiferromagnetic phase may lead to metal-insulator transition
due to change in lattice translational symmetry (supercell
symmetry), thereby splitting the Brillouin zone. The other effect
observed (e.g. in giant magnetoresistive materials) is a transition
from paramagnetic insulating phase to ferromagnetic metallic phase;
here the ferromagnetism leads to a metallic conduction. The present
case is completely opposite revealing a large band gap insulating
ferromagnetic phase in a non-magnetic metal.

\begin{figure}
\vspace{-2ex}
\begin{center}
\includegraphics[angle=0,width=0.4\textwidth]{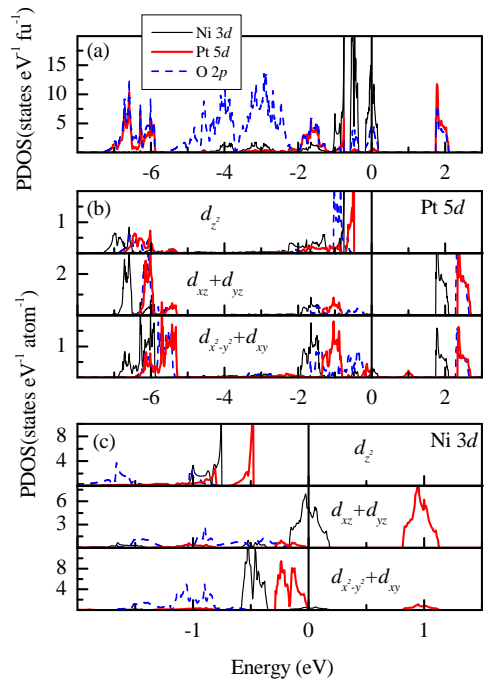}
\vspace{-4ex}
\end{center}
\caption{(Color online) (a) The partial density of states (PDOS)
corresponding to Ni 3$d$ (thin solid lines), O 2$p$ (dashed lines)
and Pt 5$d$ (thick solid lines) states for non-magnetic solution. In
(b) and (c), we show $d_{z^2}$ (upper panel), $d_{xz}+d_{yz}$
(middle panel) and $d_{x^2-y^2}+d_{xy}$ (lower panel) PDOS for Pt
5$d$ and Ni 3$d$ states, respectively. Up and down spin PDOS are
shown by dashed and thick solid lines, respectively. In each case,
half of the non-magnetic PDOS is shown by thin solid line for
comparison.
 }
 \vspace{-2ex}
\end{figure}

In order to understand the effect, we plot the character of various
energy bands in Fig. 3. The degeneracy of Ni 3$d$ in NiO$_6$
trigonal prisms and Pt 5$d$ orbitals in PtO$_6$ octahedra is lifted
due to the corresponding crystal field effect. We have defined the
axis system such that the $z$-axis lies along $c$-axis, and $x$- \&
$y$-axes are in the $ab$-plane (see Fig. 1). The crystal field
splitting of the Ni 3$d$ and Pt 5$d$ levels are evident in Fig.
3(a), where we show the Ni 3$d$, Pt 5$d$ and O 2$p$ partial density
of states (PDOS) obtained from non-magnetic solution. Pt 5$d$
dominated bonding and antibonding energy bands appear in the energy
ranges -7 to -5 eV and -2.5 to -1 eV, respectively. O 2$p$
contributions appear primarily between -2 to -5 eV. Energy bands
having dominant Ni 3$d$ character shown by thin solid line appear in
the vicinity of the Fermi level. Trigonal prismatic crystal field
around the Ni site leads to three distinct energy bands having
$d_{z^2}$, ($d_{x^2-y^2}, d_{xy}$) and ($d_{xz}, d_{yz}$) character
as seen in Figs. 3(c). Clearly, the ground state is metallic and the
electronic density of states at the Fermi level are primarily
contributed by Ni 3$d_{xz}+d_{yz}$ states.

In Fig. 3(b), we show the spin polarized Pt 5$d$ PDOS. Both the spin
channels of Pt 5$d$ states are partially occupied and the insulating
gap in the up spin channel is essentially decided by the 5$d$
states, which is found to be about 2.5 eV. From the total electron
count, we find that the Pt has 6 electrons in the 5$d$ bands; the
valence state is 4+. These 6 electrons occupy $t_{2g}$ spin-orbitals
making it completely filled. Since all these energy bands appear at
much lower energies, an exchange splitting of about 0.6 eV observed
for Pt 5$d$ states does not push the down spin bands enough towards
the Fermi level to make them partially filled. Thus, the Pt 5$d$
possess weak magnetic moment presumably induced by Ni 3$d$ moments.

Spin polarized calculations exhibit large exchange splitting (about
1 eV) in all the spin-orbitals corresponding to Ni 3$d$ electronic
states. Such a large exchange correlation does not affect the
insulating state of the $d_{z^2}$ and ($d_{x^2-y^2}, d_{xy}$) energy
bands. The spin degeneracy lifting of ($d_{xz}, d_{yz}$) bands leads
to a gap of about 0.8 eV at $\epsilon_F$. Since, Ni$^{2+}$ has 8
electrons in the $d$ band, the down spin band becomes completely
empty due to spin polarization. Thus, the Hund's coupling drives the
system towards insulating ground state.

It is to note here that capturing insulating ground state in most of
the oxides containing 3$d$ transition metals requires consideration
of on-site Coulomb interaction strength, $U$ among 3$d$ electrons.
Correlation effect is also important in higher $d$
systems.\cite{ravi,y2ir2o7} This is also the case in
Sr$_3$NiRhO$_6$.\cite{sudhir} However, in the present case, on-site
Coulomb correlation among $d$ electrons is not necessary to
determine the electronic properties of Sr$_3$NiPtO$_6$. We have
verified this by including $U$ in our calculations (LSDA+$U$); $U$
corresponding to Ni 3$d$ electrons was varied upto 7 eV and that
corresponding to Pt 4$d$ electrons upto 4 eV. We did not observe any
change in the electronic structure that will influence the
electronic and magnetic properties except an enhancement in the
energy gap with the increase in $U$ maintaining the insulating
ground state. The Ni moments are also found to be very close to the
LSDA results. This is consistent with the expectations for an
insulating system.

The total converged energy for the ferromagnetic solution is found
to be $\sim$ 781 meV/fu lower than that for the non-magnetic
solution. This clearly provides evidence against the non-magnetic
phase for this system as predicted in previous
studies.\cite{claridge} The magnetic moment centered at Ni sites is
significantly large ($\sim$ 1.5 $\mu_B$). The magnetic moment at Pt
site turns out to be negligibly small ($\sim$ 0.02 $\mu_B$)
indicating that Pt is in low spin state in the octahedral symmetry.
The total magnetic moments per formula unit is found to be about 2
$\mu_B$.

In order to investigate the effect of spin-orbit coupling (SOC) on
the magnetic states of the Ni and Pt, we performed FM GGA
calculations including SOC for both the atoms. This calculation
converges again to the insulating ground state. The spin part of the
magnetic moments do not get affected by the SOC. However, SOC
induces about 0.16 $\mu_B$ orbital magnetic moments at Ni site. The
direction of orbital and spin parts of magnetic moments are same,
which is as per the Hund's rule. The orbital part of magnetic moment
for Pt atom is found to be very small (0.02 $\mu_B$). This result is
surprising as the effect of SOC is expected to be significantly
larger for heavier Pt atom in comparison to lighter Ni atom. The
observance of negligibly small orbital magnetic moment for Pt may be
attributed to the fully filled $t_{2g}$ orbitals. The relatively
small value of orbital part of magnetic moment in comparison to spin
part clearly indicates that the magnetic properties of
Sr$_3$NiPtO$_6$ compound is primarily determined by the spin
dynamics associated to Ni 3$d$ electrons.

It is already well known that magnetic interactions can be captured
well via {\it ab initio} calculations.\cite{kbm} In order to learn
the nature of magnetic interaction among Ni atoms, we calculated the
electronic structure for antiferromagnetic coupling among
neighboring intra-chain Ni atoms. The energy of this solution is
found to be about 5 meV/fu less than that for the FM solution
indicating that antiferromagnetic interaction provides more
stability to the system, which is consistent with magnetization
data.\cite{niharika} The calculation gives almost the same magnetic
moments ($\sim$ 1.5 $\mu_B$) as observed before; the net magnetic
moment is close to zero. This Ni moment is very close to that found
in Sr$_3$NiRhO$_6$.\cite{sudhir}

\begin{figure}
\vspace{-2ex}
\begin{center}
\includegraphics[angle=0,width=0.4\textwidth]{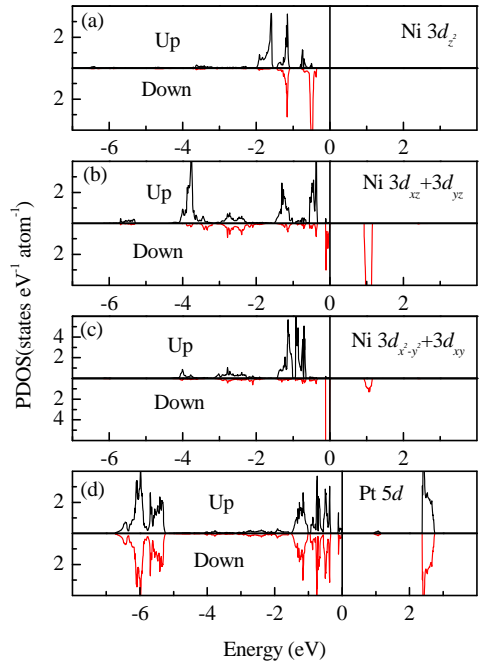}
\vspace{-4ex}
\end{center}
\caption{(Color online) Up and down spin PDOS corresponding to (a)
Ni3$d_{z^2}$ (b) Ni3$d_{xz}+d_{yz}$, (c) Ni 3$d_{x^2-y^2}+d_{xy}$,
and Pt 5$d$ states for antiferromagnetic solution.
 }
 \vspace{-2ex}
\end{figure}

Now we discuss the effect of antiferromagnetic interaction on the
electronic structure of the compound. Since the energy distribution
of the density of states for both the nonequivalent Ni sites (due to
antiparallel ordering) are almost identical, we plot the partial
density of states corresponding to only one Ni site in Fig. 4. The
3$d$ orbitals having different symmetries are plotted in separate
panels of the figure for clarity. The density of states clearly
manifest again the insulating phase. The band gap of about 1 eV is
observed in the down spin channel as in the ferromagnetic case and
the gap in up spin channel is much higher. The Pt 3$d$ PDOS appear
within -1.7 eV from the Fermi level. The small exchange splitting
observed in ferromagnetic solution is now absent. This is expected
as each Pt has one up-spin and one down-spin Ni neighbor. Therefore,
the induced moment due to finite orbital overlaps will be vanished.
Similar behavior is also seen for O 2$p$ PDOS.

\begin{figure}
\vspace{-2ex}
\begin{center}
\includegraphics[angle=0,width=0.4\textwidth]{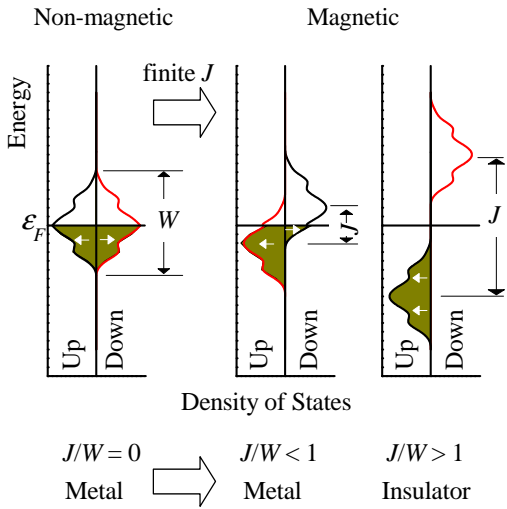}
\vspace{-24ex}
\end{center}
\caption{A demonstration of how an insulating phase appears due to
finite exchange coupling strength, $J$. For $J/W >$ 1 ($W$ =
bandwidth) the system becomes insulating and $J/W <$~1 it remains
metallic.
 }
 \vspace{-2ex}
\end{figure}

In summary, we show an example of emergence of an insulating phase
from a metallic  phase due to finite exchange interaction strength.
The metal to insulator transition, here, is driven by exchange
splitting and independent of magnetic coupling among moments located
at different sites. This is demonstrated schematically in Fig. 5.
E.g. in a simple model consisting of half filled bands, if the
exchange splitting $J$ is larger than the bandwidth $W$ ($J/W
>$~1) the system becomes insulating purely due to the exchange
interactions. $J/W <$~1 corresponds to metallic ground state. Such a
metal-insulator transition can be achieved experimentally by tuning
the bandwidth as often done in correlated electron systems via
change in bond-angles and/or exchange interaction strength via
chemical substitutions.

The Ni moments are found to be large in Sr$_3$NiPtO$_6$.
Consideration of electron correlation appears to be not necessary to
determine the electronic properties and magnetic moments. The {\em
ab initio} results reveal that Ni moments are coupled
antiferromagnetically along the chain. As the inter-chain
interactions are complicated ($M$-O-O-$M$ super superexchange
interaction), intra-chain interactions become important and probably
play the key role for the spin liquid phase.


The authors acknowledge Prof. E.V. Sampathkumaran, TIFR, India for
drawing our attention towards this compound and useful discussions.

\end{document}